\documentclass[aps,preprint,amsmath,amssymb,showpacs]{revtex4}

\usepackage{graphicx}
\usepackage{dcolumn}
\usepackage{bm}

\newcommand {\be}{\begin{equation}} 
\newcommand {\ee}{\end{equation}}

\def \be{\begin{equation}}
\def \ee{\end{equation}}
\def \bea{\begin{eqnarray}}
\def \eea{\end{eqnarray}}

\def \ebar{\overline{\epsilon}}

\def \Qu{Q_{\mbox{\tiny U}}}

\def \ebar{\overline{\epsilon}}

\def \lbar{\overline{\ell}}

\def \d{\delta}

\def \Qf{Q_{\mbox{\tiny F}}}

\begin{document}

\title{Protein folding rates correlate with heterogeneity of folding mechanism}

\author{B. \"Oztop$^{\ddag}$}
\author{M. R. Ejtehadi$^{\ddag,\dag}$}
\author{S. S. Plotkin$^{\ddag}$} 
\email{steve@physics.ubc.ca}
 
\affiliation{${}^\ddag$ Department of Physics and Astronomy, 
University of British Columbia, 
Vancouver, BC V6T-1Z1, 
Canada\\
${}^\dag$ Department of Physics, 
Sharif University of Technology, 
Tehran 11365-9161, Iran
}

\date{\today}

\begin{abstract}
By observing trends in the folding kinetics of experimental 2-state proteins at 
their transition midpoints, and by observing trends in the barrier heights of 
numerous simulations of coarse grained, C$_{\alpha}$ model, G\={o} proteins, 
we show that folding rates correlate with the degree of heterogeneity in the 
formation of native contacts. Statistically significant correlations are 
observed between folding rates and measures of heterogeneity inherent in 
the native topology, as well as between rates and the variance in the 
distribution of either experimentally measured or simulated $\phi$-values. 
\end{abstract}


\maketitle

Protein folding is a relaxation process driven by a first order like
fluctuation of a critical
nucleus~\cite{note:fluctuation-nucleus}.
Because proteins are 
evolutionarily designed to fold to a particular structure, frustrating
interactions are minimized and the folding process can be projected
onto one or a few reaction coordinates without too much loss of
information~\cite{note:projection-reaction-coord}.
This projection yields a
free energy surface whose  
structure is subject to much interest. Different proteins have
different free energy surfaces with different barrier heights.

What factors determine the height of the folding free energy barrier
for the various proteins? As one would expect, the barrier decreases
as the energetic stability of the folded structure
increases~\cite{DinnerAR01}.  Moreover folding rates tend to increase
with energetic discrimination measures between the folded state and
unfolded or misfolded decoys~\cite{MelinR99}.  As one might also
expect, the barrier increases for native structures that have longer
polymer loops formed during folding.  A
property capturing this effect, dubbed absolute contact order (ACO),
measures the mean sequence separation between amino acids in close
proximity (and thus interacting) in the native
structure~\cite{note:topology-note}: $ ACO \equiv \lbar = (1/M)
\sum_{i<j} |i-j| \Delta_{ij}^N $ where $i$ and $j$ label amino acid index,
$\Delta_{ij}^N = 1$ (or 0) if amino acids $i$ and $j$ are (or are not)
interacting in the native structure, and M is the total number of
contacts in the native structure determined by either heavy side chain
atoms or C$_{\alpha}$ atoms within a cut-off distance of 4.8 \AA
~\cite{note:off-latt-model}.

In what follows we first re-examine the trend of rates with $\lbar$ in
light of theoretical predictions~\cite{note:funct,PlotkinSS00:pnas,PlotkinSS02:Tjcp}, 
then we will go on to further examine higher-order aspects of native topology 
(and energetics) that act as predictors of folding rate. 

\begin{figure}
\includegraphics[height=10cm]{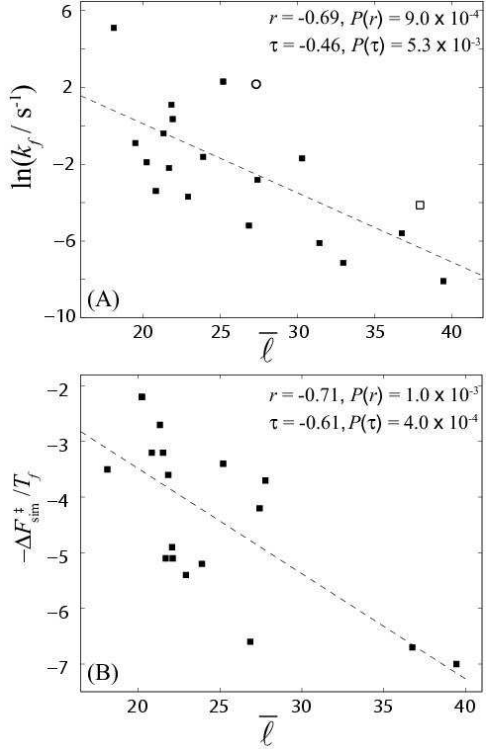}
\caption{\label{fig:aco}
(A) Logarithm of experimental folding rate (in $sec^{-1}$) at the transition midpoint 
$vs$ absolute contact order or mean sequence separation between interacting residues 
in the native structure, $\lbar$. Wild type protein S6 is shown by an open square and 
P$^{13-14}$ circular permutant of S6~\cite{LindbergM02} is shown by an open circle. 
(B) The equivalent 
measure in G\={o} simulations is $-\Delta F_{sim}^{\ddag} / T_f$, again plotted $vs$ 
$\lbar$. Both show a statistically significant anti-correlation: $r$ (or $\tau$) is the 
correlation coefficient (or Kendall's Tau). Statistical significance is defined here by 
the probability $P(r)$ ($P(\tau)$) to observe a given correlation coefficient or greater 
by chance. If $P(r)$ ($P(\tau)$) $<$ 0.05, the dependence is typically deemed 
statistically significant~\cite{note:stat-sig}. 
Shown in (A) are 19 proteins (and P$^{13-14}$ circular permutant of S6) 
for which experimental rate data are available 
at various denaturant concentrations~\cite{note:prots} and in (B) 18 simulated G\={o} 
model proteins~\cite{note:simPDB}.
}
\end{figure}

If we take data that first corrects for the effects of differing native stabilities
for different proteins by
adjusting denaturant concentration to conditions at the transition
midpoint, and then plot the log folding rate $vs$ $\lbar$, we
find a statistically significant correlation for a representative set
of 19 2-state proteins (and P$^{13-14}$ circular permutant of S6)
(Fig.~\ref{fig:aco}A)~\cite{note:prots}. Observations similar to this led the
folding community to accept the idea that properties of native
topology strongly determine folding rate~\cite{note:topology-rate}. 
Moreover if one simulates off-lattice
C$_{\alpha}$ G\={o} models~\cite{note:off-latt-model} to 18 structures
of known 2-state folders~\cite{note:simPDB}, one also finds a
statistically significant correlation between barrier height and
absolute contact order (Fig.~\ref{fig:aco}B). One also notices from Fig.~\ref{fig:aco}
that there must be more to the story then absolute contact order in
determining folding rates, since the fluctuations around the best fit
line are significant.

The effects of native topology (and energetics) should be describable
analytically as well. To this end a free energy functional approach
was developed~\cite{note:funct,PlotkinSS00:pnas,PlotkinSS02:Tjcp}
within which it was shown that the
free energy barrier may be written in terms of an expansion involving
moments of distributions of native contact interaction energies
$\{\epsilon_{ij}\}$, and native contact sequence separations $\{\ell_{ij}\} \equiv
\{|i-j|\}$. The lowest order corrections to the mean-field barrier
are~\cite{PlotkinSS02:Tjcp}:  
\be
\frac{\Delta F^{\ddag}}{MT}(\{\epsilon_{ij}\},\{\ell _{ij}\})   =
\frac{\overline{\Delta F}^{\ddag}}{M T}
- \! A 
\frac{\overline{\delta\epsilon^2}}{ T^2} - \! B
\frac{\overline{\delta\epsilon \, \delta \ell}}{\lbar \, T} - \! C 
\frac{\overline{\delta \ell^2}}{\lbar^2}
\label{eq:barrier}
\ee
where $A$, $B$, $C$ are all positive and of order unity. The
lowest order mean field term $\overline{\Delta F}^{\ddag} \equiv
\Delta F^{\ddag} (\ebar, \lbar)$, where $\overline{\epsilon}$, $\lbar$ are 
the first moments (mean) of the distributions, indeed   
increases as $\lbar$ increases, consistent with the observed trend. 
The theory gives the slope $m_{\mbox{\tiny{MF}}}$ of the mean field barrier $vs$ $\lbar$
as~\cite{PlotkinSS02:Tjcp} 
\begin{equation}
m_{\mbox{\tiny{MF}}} \equiv \partial (-\overline{\Delta F}^{\ddag}/T) \mbox{\Large /} \partial \lbar
\approx -(3/2) (M / \lbar^2) \ln(\lbar^{1/2}/2) .
\label{eq:slope}
\end{equation}

Calculating Eq.~(\ref{eq:slope}) for all proteins used in Fig.~\ref{fig:aco}A, 
$m_{MF} = -0.41 \pm 0.09$,
which is consistent with the slope of the best fit line $-0.36$. The mean field slope 
for the proteins in Fig.~\ref{fig:aco}B is $-0.42\pm0.08$ which is almost twice the 
slope of the best fit line $-0.19$. There may be several reasons for this, including the fact
that the theory used the mean field approximation, while the nucleus may be better 
approximated by a capillary model, and the Gaussian approximation for polymer loops 
used in the theory may be poor for many contacts. There may also be a cancellation of errors in 
Fig.~\ref{fig:aco}A due to the presence of a capillary nucleus with many-body interactions 
present~\cite{note:EAP}, which would result in unexpectedly good agreement. 

Second order terms in Eq.~(\ref{eq:barrier}) involving the fluctuations of
native energies and loop lengths contact to contact all tend to
decrease the barrier, leading to the notion that proteins with more
heterogeneous folding mechanisms should fold
faster~\cite{PlotkinSS00:pnas,PlotkinSS02:Tjcp}. We note that here a more heterogeneous
folding mechanism corresponds to a more specific, polarized folding
nucleus, i.e. the heterogeneity here refers to contact formation probability,
not conformational diversity of the transition state.  Earlier
lattice-simulation studies~\cite{AbkevichVI94} as well as more recent
experimental studies of circular permutants~\cite{LindbergM02} support
the notion that a more polarized nucleus results in a faster folding
protein.

We can readily check if the second moment of the loop length
distribution has an observable effect on rates, even if we ignore 
variations due to different $\lbar$ values protein to protein, as well as
the
terms with coefficients $A$ and $B$ in Eq.~(\ref{eq:barrier}).
The functional theory gives coefficient $C \approx Q^{\ddag}$ in 
Eq.~(\ref{eq:barrier}) ~\cite{PlotkinSS02:Tjcp}, 
so the change in barrier height due to the presence of structural variance is:
\begin{equation}
(\Delta F^{\ddag} - \overline{\Delta F}^{\ddag}) /M T \equiv
\delta\Delta F^{\ddag}/M T \approx -Q^{\ddag} \overline{\delta
  \ell^2}/\lbar^2 .
\label{eq:DFstructdisp}
\end{equation}
Here, $Q$ is the overall fraction of native contacts, and $Q^{\ddag}$ is the
value of $Q$ at the barrier peak. 

\begin{figure}
\includegraphics[height=10cm]{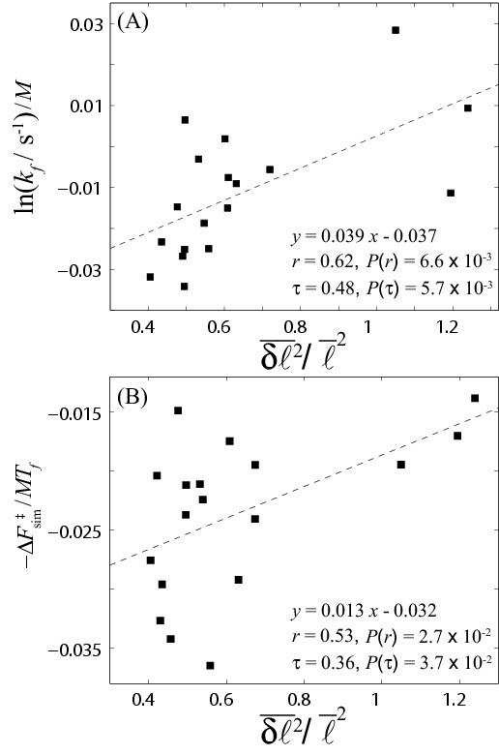}
\caption{\label{fig:str} Plotted in (A) are log experimental rate data
  (at the transition midpoints) and in (B),  
simulated barriers (at $T_f$), as a function of the measure of
  structural heterogeneity that appears in the functional theory in
  Eq.s~(\ref{eq:barrier}) and~(\ref{eq:DFstructdisp}).
 Both show a moderate, but statistically significant 
correlation with structural variance. Three $\alpha/\beta$ proteins ($\lambda$-repressor 
chain 3, cytochrome c, yeast iso-1-cytochrome c) tend to have both large structural 
variance and fast folding rates.
}

\end{figure}

Plots of experimental log folding rate and simulated barrier
heights (over $M T$) both show statistically significant correlation with
$\overline{\delta \ell^2} / \lbar^2$ (Fig.~\ref{fig:str}).

However there are large fluctuations present, and the slope of the best fit line 
is only about a tenth the theoretical prediction. Neglecting trends due to contact 
order and energetic variance introduces errors in the plots. 

Experimentally measured $\phi$-values~\cite{FershtAR99:book} involve both energetics and 
entropics and should better capture the effects of heterogeneity in folding mechanism. 
The variance in $\phi$-values couples together the last 3 terms in Eq.~(\ref{eq:barrier}). 
To facilitate a comparison of rates with $\phi$-variance, the free energy barrier maybe 
recast in terms of the variance in native contact formation
probabilities ($Q_{ij}$)~\cite{PlotkinSS02:Tjcp} 
\begin{equation}
\delta\Delta F^{\ddag}/M T 
\approx -  \overline{\delta Q^2}/2 Q^{\ddag} .
\label{eq:barrier-Q}
\end{equation}

Eq.~(\ref{eq:barrier-Q}) only includes the effects of heterogeneity in
polymer loop length, however energetic heterogeneity can be
incorporated as well, which only changes the coefficient 
$(1/2 Q^{\ddag})$ in Eq.~(\ref{eq:barrier-Q}) to $(3/2 Q^{\ddag})$. 
The simulations have no variance in native contact energies, moreover 
statistics arguments suggest that this native variance may be significantly 
reduced with respect to the variance in collapsed random 
structures~\cite{note:projection-reaction-coord}.

\begin{figure}
\includegraphics[height=10cm]{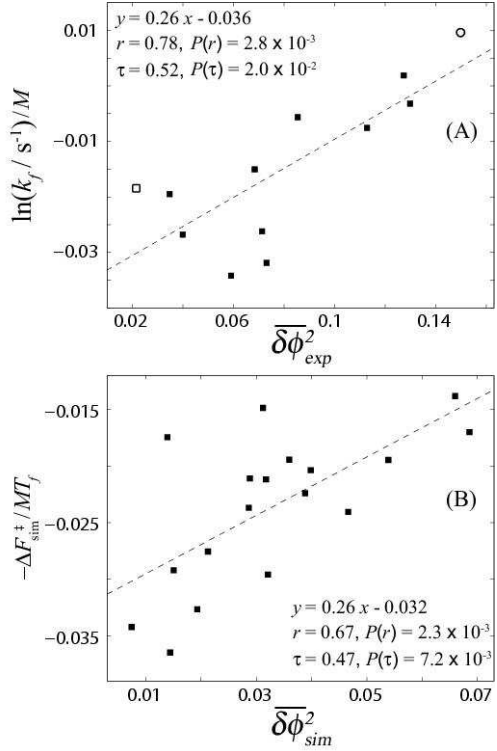}
\caption{\label{fig:phi} Plots of log experimental folding rate (over
$M$) for a subset of proteins in Fig.~\ref{fig:aco}A for which experimental 
$\phi$ values are available and minus 
free energy barrier (over $MT$) for simulated proteins $vs$
$\phi$-variance. Both show strong statistically significant
correlation. In particular the trend in experimental data is strong
even though the number of proteins with available data for both
$\phi$-variance and transition midpoint rate is not large. Experimental 
data for wild type S6 is shown by an open square and P$^{13-14}$ circular 
permutant of S6~\cite{LindbergM02} is shown by an open circle which 
fits very well to the rest of the data and increases correlation. 
The strong correlation remains upon dividing by chain length $N$ instead 
of total number of contacts $M$.}
\end{figure}

$\phi$-values may be defined analytically as~\cite{Onuchic96,note:EAP}
\begin{equation}
\phi_i = \displaystyle\sum_{j\neq i} (Q_{ij}^{\ddag} -
  Q_{ij}^{\mbox{\tiny U}} ) \Delta_{ij}^N \mbox{\LARGE{/}} \displaystyle\sum_{j\neq i} 
(Q_{ij}^{\mbox{\tiny F}}
  - Q_{ij}^{\mbox{\tiny U}} ) \Delta_{ij}^N 
\label{eq:phidef}
\end{equation}
where $Q_{ij}^{\mbox{\tiny U}}$, $Q_{ij}^{\ddag}$ and $Q_{ij}^{\mbox{\tiny F}}$ are the
probabilities of native contact formation between residues $i$ and $j$ in the
unfolded, transition and folded states respectively.  

In the approximation that all contacts are fully formed in the native
structure ($\Qf=1$), and unformed in the unfolded structures ($\Qu =
0$), the $\phi$-value for residue $i$ is the mean of $Q_{ij}$ values in the
transition state (c.f. Eq.~(\ref{eq:phidef})). Further approximating the same number of nearest
neighbors $z$  for all residues, the variances are related by
$\overline{\d \phi^2} \approx (1/z) \overline{\d Q^2}$. If we make no
approximations and simply plot $\overline{\d Q^2}$ {\it vs.}
$\overline{\d \phi^2}$ (for the simulation data), the quantities correlate extremely well
(see Table I) with a slope of $\approx 1.2$ and an intercept
$-0.04$ . The intercept may
be non-zero since other fluctuating quantities (e.g. $\Qu$, $z$)
contribute to the variance of $\phi$-values. 

The above arguments indicate $\overline{\d Q^2}$ and $\overline{\d \phi^2}$
are within a factor of approximately unity, 
so we rewrite Eq.~(\ref{eq:barrier-Q}) 
in the form 
\begin{equation}
\delta\Delta F^{\ddag} / M T \approx -D \: \overline{\delta\phi^2}
\label{eq:phi-bar}
\end{equation}
with $D$ a parameter of order unity.

According to Eq.~(\ref{eq:phi-bar}) more polarized nuclei have lower free energy 
barriers. Plots of $-\Delta F^{\ddag}/M T$ $vs$ $\overline{\delta\phi^2}$ for experiments 
and simulations are shown in Fig.~\ref{fig:phi}. Here we see a strong statistically 
significant correlation of both rates and barriers with $\phi$ variance. Moreover 
the slopes of the best fit lines ($\approx 0.3$) compare somewhat more favorably with the
theoretically predicted values ($\approx 0.8$) than was the case for
structural variance. A precise comparison with experimental data is
more difficult since the coordination number $z$ as well as the
numbers $\Qu$ and $\Qf$ are  not accurately known for all proteins. Taking
the slope from Fig.~\ref{fig:phi}A and using the approximations
mentioned above allows us to infer the residue-residue coordination
number: $z \approx 4$ if energetic
heterogeneity is negligible (Eq.~(\ref{eq:barrier-Q})), $z \approx 11$ if it
is substantial (Eq.~(\ref{eq:barrier-Q}) with coefficient $3/2 Q^{\ddag}$).

The residuals of $-\Delta F^{\ddag} / M T$ $vs$ $\lbar$, when plotted against 
$\overline{\delta \ell^2}/\lbar^2$ and $\overline{\delta \phi^2}$, show 
comparable but typically slightly less significant correlation (within 10$\%$)  
to those in Fig.~\ref{fig:str} and Fig.~\ref{fig:phi}. The term 
$\delta\Delta F^{\ddag} / M T$ can be thought of as a measure of these
residuals. We have plotted absolute rates, which are easily measurable from
experiments or simulations, while the mean-field barrier is not. 

We note that $\overline{\delta\phi^2}_{\!\!\! exp}$ has errors due
both to experimental measurement as well as the small set of $\phi$-values for
each protein. Moreover the experimental rates at the transition
midpoint are compared to the variance in $\phi$'s typically measured
in water or stabilizing conditions. Interestingly, experimental
folding mechanisms tend to be more polarized than uniform G\={o}
models.

In the case of the simulations, the correlation between $\overline{\delta\phi^2}$ $vs$
$\overline{\delta \ell^2}/\lbar^2$ is strong as expected, 
since there is no variance in native contact energies, by construction
of the model. 
For experimental data however the correlation is poor, which implies that
there may be substantial energetic heterogeneity present in native contact
energies of real proteins. 
It is not too surprising then that there is no correlation
between the variance of experimental $\phi$-values and simulation
$\phi$-values (see Table~\ref{tab:correlations}). 
In the analysis then, simulated barriers were plotted against
simulated $\phi$-variance, and experimental rates were plotted against
experimental $\phi$-variance.

We did not find any significant correlation between rates and
structural variance $\overline{\d \ell^2}/\lbar^2$ for 3-state
folders. Here there is the intriguing picture that (on-pathway)
intermediates in 3-state folders are in fact induced by structural or
energetic heterogeneity, so that there is no {\it a priori} reason for
folding rates to continue to increase with increasing heterogeneity. 

S6 displays significant correlation between native contact energies
and native loop lengths~\cite{LindbergM02}. For this reason we did not
include it in Fig.~\ref{fig:str}A, which only includes a structural
measure of heterogeneity- if it is included the correlation decreases
to $r=0.57$, $P(r)=9.6\times 10^{-3}$. 
We note that the inclusion of the two data points corresponding to 
S6 does not change the
correlation in Fig.~\ref{fig:phi}A  and {\it decreases} the
correlation in Fig.~\ref{fig:aco}A by $8\%$.

We showed here that both experimental rates and simulated free energy
barriers for 2-state proteins depend on
the degree of heterogeneity present in the folding process.  The
results compared quite well with the predictions of the free energy
functional theory~\cite{PlotkinSS00:pnas,PlotkinSS02:Tjcp}.
Heterogeneity due to variance in the distribution of native loop
lengths, as well as variance in the distribution of $\phi$-values,
were both seen to increase folding rates and reduce folding barriers.
The observed effect due $\phi$-variance was the most statistically significant (as
expected), because $\phi$-variance captures both heterogeneity arising from native
topology as well as that arising from energetics.

\begin{acknowledgments}
S.~S.~P. acknowledges support from the Natural Sciences and Engineering 
Research Council and the Canada Research Chairs program. We thank 
Kevin Plaxco and Mikael Oliveberg for helpful discussions.
\end{acknowledgments}

\begin{table}
\caption{\label{tab:correlations} Correlation coefficient and statistical significance for various quantities.}
\begin{ruledtabular}
\begin{tabular}{llllll}
$y$\hspace{.8cm} $vs$ & $x$ & $r$ & $P(r)$\footnotemark[1] & $\tau$ & $P(\tau)$\footnotemark[1]\\
\hline
$\ln (k_f)$ & $\lbar$ & -0.69 & 9$\times 10^{-4}$ & -0.46 & 5.3$\times 10^{-3}$\\
$-\Delta F_{sim}^{\ddag}/T_f$ & $\lbar$ & -0.71 & $10^{-3}$ & -0.61 & 4$\times 10^{-4}$\\
$\ln (k_f)/M$\footnotemark[2] & $\overline{\delta\phi^2}_{\!\!\! exp}$ & 0.78 & 2.8$\times 10^{-3}$ & 0.52 & 2$\times 10^{-2}$\\
$-\Delta F_{sim}^{\ddag}/M T_f$ & $\overline{\delta\phi^2}_{\!\!\! sim}$ & 0.67 & 2.3$\times 10^{-3}$ & 0.47 & 7.2$\times 10^{-3}$\\
$\ln (k_f)/M$ & $\overline{\delta\ell^2}/\lbar^2$ & 0.62 & 6.6$\times 10^{-3}$ & 0.48 & 5.7$\times 10^{-3}$\\
$-\Delta F_{sim}^{\ddag}/M T_f$ & $\overline{\delta\ell^2}/\lbar^2$ & 0.53 & 2.7$\times 10^{-2}$ & 0.36 & 3.7$\times 10 ^{-2}$\\
\hline
$\lbar$ & $\overline{\delta\ell^2}/\lbar^2$\footnotemark[3] & -0.14 & 0.52 & -0.07 & 0.7\\
$\lbar$ & $\overline{\delta\phi^2}_{\!\!\! exp}$ & -0.64 & 2.5$\times 10^{-2}$ & -0.43 & 5.5$\times 10^{-2}$\\
$\lbar$ & $\overline{\delta\phi^2}_{\!\!\! sim}$ & 0.16 & 0.52 & 0.15 & 0.38\\
$\overline{\delta\phi^2}_{\!\!\! sim}$ & $\overline{\delta\ell^2}/\lbar^2$ & 0.71 & $10^{-3}$ & 0.32 & 6.4$\times 10^{-2}$\\
$\overline{\delta\phi^2}_{\!\!\! exp}$ & $\overline{\delta\ell^2}/\lbar^2$ & 0.29 & 0.37 & 0.18 & 0.41\\
$\overline{\delta\phi^2}_{\!\!\! exp}$ & $\overline{\delta\phi^2}_{\!\!\! sim}$ & -0.16 & 0.8 & 0.2 & 0.63\\
$\overline{\delta\phi^2}_{\!\!\! sim}$ & $\overline{\delta Q^2}_{\!\!\!sim}$ & 0.94 & $< 10^{-6}$ & 0.77 & 9$\times 10^{-6}$\\

\end{tabular}
\end{ruledtabular}
\footnotetext[1]{2-sided statistical significance has been used.}
\footnotetext[2]{Here we divide by the number of native contacts $M$. 
Dividing instead by chain length $N$ gives correlations within 10$\%$.
$M$ and $N$ correlate very strongly ($r = 0.94$).}
\footnotetext[3]{Data from both simulated and experimental proteins used.}
\end{table}

\end{document}